\documentclass[aps,prb,reprint,superscriptaddress]{revtex4-2}
\UseRawInputEncoding
\usepackage{amsmath,amssymb}
\usepackage{graphicx}
\usepackage[colorlinks,linkcolor=blue,anchorcolor=blue,citecolor=blue,urlcolor=blue]{hyperref}
\usepackage{booktabs}
\usepackage{multirow}
\usepackage[textsize=tiny,backgroundcolor=yellow]{todonotes}

\begin{document}


\title{Kinetic instability and superconductivity in Li$_2$AuH$_6$ and Li$_2$AgH$_6$ at ambient pressure}



\author{Yucheng Ding}
\affiliation{International Center for Quantum Materials, Peking University, Beijing 100871, China}
\author{Haoran Chen}
\affiliation{International Center for Quantum Materials, Peking University, Beijing 100871, China}
\author{Junren Shi}
\email{junrenshi@pku.edu.cn}
\affiliation{International Center for Quantum Materials, Peking University, Beijing 100871, China}
\affiliation{Collaborative Innovation Center of Quantum Matter, Beijing 100871, China}


\date{\today}

\begin{abstract}
Li$_2$AuH$_6$ and Li$_2$AgH$_6$ have been proposed as promising candidates for high-temperature superconductors under ambient pressure. While previous studies confirm the dynamic stability of these two thermodynamically unstable systems, their kinetic stability against quantum and thermal fluctuations remains to be verified. In this work, we use path integral molecular dynamics simulations to examine the kinetic stability of Li$_2$AuH$_6$ and Li$_2$AgH$_6$ under ambient pressure. We find both compounds are kinetically unstable. Li$_2$AgH$_6$ undergoes lattice collapse, whereas Li$_2$AuH$_6$ retains a stable fluorite-type Li-Au sublattice, but hydrogen atoms partially dimerize into molecules and diffuse within the host lattice. Using the stochastic path-integral approach, which is a nonperturbative approach applicable to systems with diffusive atoms, we investigate the superconductivity of Li$_2$AuH$_6$ in this state. We predict a superconducting transition temperature of 22~K, well below earlier predictions, due to the low density of states at the Fermi level caused by the collapse of hydrogen sublattice and hydrogen dimerization.
\end{abstract}


\maketitle

\section{Introduction}
Ever since the discovery of superconductivity in mercury, people have been searching for superconductors with higher superconducting transition temperatures ($T_c$). According to the Bardeen-Cooper-Schrieffer (BCS) theory, hydrides are excellent candidates for achieving high $T_c$ because the light hydrogen atoms provide both high phonon frequency and strong electron-phonon coupling (EPC). Over the past decade, various high-$T_c$ hydride superconductors have been theoretically predicted and experimentally synthesized~\cite{hydride_review}, such as H$_3$S~\cite{H3S_exp}, LaH$_{10}$~\cite{LaH10_exp1,LaH10_exp2}, CaH$_6$~\cite{CaH6_exp1,CaH6_exp2}, YH$_6$ and YH$_9$~\cite{YH6_exp1,YH_exp2,YH_exp3}, all exhibiting $T_c$ exceeding 200~K, with recent progress in LaSc$_2$H$_{24}$~\cite{LaSc2H24} even approaching room-temperature superconductivity. However, these high-$T_c$ hydrides are only stable under extremely high pressure, preventing them from practical applications.

Given this, in the past few years, attention has shifted to high-$T_c$ hydride superconductors under ambient pressure. In 2024, Dolui \textit{et al.} and Sanna \textit{et al.} independently proposed Mg$_2$IrH$_6$ as an ambient-pressure hydride superconductor, with a maximum predicted $T_c$ of 160~K~\cite{Mg2IrH6,X2MH6_1}. Subsequent studies revise this prediction downward and propose other candidates with the same X$_2$MH$_6$ crystal structure, which is known as the SM$_2$-TM-H$_6$ family~\cite{X2MH6_1,X2MH6_2,ambient_search}. These studies also identify additional candidates with different crystal structures~\cite{ambient_search,MXH3,MH4,HC6,AXH8,ABH6,A2XBH6,ambient_search_2}. Among them, Li$_2$AuH$_6$, proposed by Ouyang \textit{et al.}, is a member of the SM$_2$-TM-H$_6$ family, with a predicted $T_c$ of 140~K~\cite{Li2AuH6}. Later, Gao \textit{et al.} conduct an extensive structure search and identify Li$_2$AgH$_6$ and Li$_2$AuH$_6$ as candidates that likely achieve the highest $T_c$ in their dataset, with predicted $T_{c}$ values ranging from 80 to 120~K~\cite{ambient_maximum}. It is further predicted that the $T_c$ of Li$_2$AuH$_6$ may be enhanced by applying a moderate pressure~\cite{Li2AuH6_pressure,X2MH6_elph}.

Structural stability has long been a major concern for high-$T_c$ hydride superconductors under ambient pressure~\cite{ambient_search,ambient_search_2}. Candidates with higher predicted $T_c$ values tend to be increasingly thermodynamically unstable~\cite{ambient_maximum,ambient_thermodynamic}. Consequently, they are necessarily metastable, provided they possess both dynamic and kinetic stability~\cite{Mg2IrH6}. It should be noted that dynamic stability is not equivalent to kinetic stability. While dynamic stability indicates stability against small perturbations in atomic positions and is examined via phonon dispersion calculations, kinetic stability requires stability against structural transitions at finite temperature and is usually assessed using molecular dynamics (MD).

In previous studies, Li$_2$AuH$_6$ and Li$_2$AgH$_6$ are considered thermodynamically unstable yet dynamically stable~\cite{Li2AuH6,ambient_maximum}, but their kinetic stability remains questionable. Their dynamic stability is supported by the absence of imaginary modes in their phonon dispersions within the stochastic self-consistent harmonic approximation (SSCHA)~\cite{SSCHA_1,SSCHA_2,SSCHA_3,SSCHA_4}. In principle, SSCHA incorporates both nuclear quantum and thermal anharmonic fluctuations within a variational framework for the free energy, which allows for the assessment of kinetic stability. In practice, however, SSCHA relies on a Gaussian ansatz for the many-body density matrix that assumes the existence of atomic equilibrium positions and effective phonons~\cite{SSCHA_4}. This ansatz may break down for systems with strong atomic diffusion. On the other hand, path integral molecular dynamics (PIMD)~\cite{PIMD,AI_PIMD} serves as a more reliable approach for evaluating kinetic stability. It accounts for both ionic quantum and thermal fluctuations without relying on the aforementioned Gaussian ansatz and is particularly suitable for hydrides, as the quantum effects of hydrogen atoms are often deemed important~\cite{H_PIMD,LaH10_SSCHA,LMH_SPIA,LaH10_SPIA}.

In this paper, we investigate the kinetic stability and superconductivity of Li$_2$AuH$_6$ and Li$_2$AgH$_6$ at ambient pressure. By performing \textit{ab initio} MD and PIMD simulations for Li$_2$AuH$_6$ and Li$_2$AgH$_6$ at 80~K, we find both of them to be kinetically unstable. Li$_2$AgH$_6$ undergoes lattice collapse, whereas Li$_2$AuH$_6$ retains a stable fluorite-type Li-Au sublattice, but hydrogen atoms partially dimerize into molecules and diffuse within the host lattice. Additionally, we investigate the superconductivity of Li$_2$AuH$_6$ in this state using the stochastic path-integral approach (SPIA)~\cite{SPIA,H_SPIA,H3S_SPIA}, which is a nonperturbative approach applicable to systems with diffusive atoms. We predict a $T_c$ well below 80~K due to the low density of states (DOS) at the Fermi level caused by the collapse of hydrogen sublattice and hydrogen dimerization.

The remainder of the paper is organized as follows. In Sec.~\ref{sec:kinetic}, we study the kinetic stability of Li$_2$AuH$_6$ and Li$_2$AgH$_6$ under ambient pressures using MD and PIMD simulations. The computational details are provided in Sec.~\ref{sec:PIMD_computation} and the results are shown in Sec.~\ref{sec:PIMD_result}. In Sec.~\ref{sec:SC}, we apply SPIA to study the superconductivity of Li$_2$AuH$_6$. We give a brief introduction of SPIA in Sec.~\ref{sec:SPIA} and present the computational details and results in Sec.~\ref{sec:SPIA_computation} and Sec.~\ref{sec:SPIA_result}. Finally, we summarize our results in Sec.~\ref{sec:summary}.

\section{Kinetic instability of L\lowercase{i}$_2$A\lowercase{u}H$_6$ and L\lowercase{i}$_2$A\lowercase{g}H$_6$\label{sec:kinetic}}
\subsection{Computational details\label{sec:PIMD_computation}}
The crystal structures of Li$_2$AuH$_6$ and Li$_2$AgH$_6$ of $Fm\bar{3}m$ space group at ambient pressure are taken from Refs.~\cite{Li2AuH6,ambient_maximum,Li2AuH6_pressure} and structurally optimized, with lattice parameters of the primitive cell being 4.71~\AA\ and 4.62~\AA, respectively. All MD, PIMD and density functional theory (DFT) calculations are performed using a modified version of the Vienna \textit{ab initio} simulation package (VASP) code~\cite{VASP,H_SPIA}. The projector-augmented wave (PAW) method~\cite{PAW} is used to describe the ion-electron interaction, and the Perdew-Burke-Ernzerhof (PBE) functional~\cite{PBE} is used to describe the exchange-correlation effect of electrons.

All simulations are performed in the canonical (NVT) ensemble at 80~K with a Langevin thermostat to control the temperature~\cite{Langevin}. The friction coefficient of the Langevin thermostat is set to 10~ps$^{-1}$ for all atomic centroid modes, and a time step of 0.5~fs is used. In DFT calculations, we use an energy cutoff of 400~eV for plane waves to expand electron wave functions. In PIMD simulations, the imaginary time is discretized with bead number $N_b=16$. 

First, we perform preliminary MD and PIMD simulations in a small $2\times2\times2$ supercell containing 72 atoms with a short simulation time of 1~ps to assess the kinetic stability of Li$_2$AuH$_6$ and Li$_2$AgH$_6$, where a $3\times3\times3$ $\Gamma$-centered $\boldsymbol{k}$-point grid is used to sample the Brillouin zone of the supercell. For more accurate analysis of kinetic stability and superconductivity, we then perform PIMD simulations for Li$_2$AuH$_6$ in a larger $3\times3\times3$ supercell containing 243 atoms with a longer simulation time of 6~ps, where a $2\times2\times2$ $\Gamma$-centered $\boldsymbol{k}$-point grid is used in this case.

\subsection{MD and PIMD results\label{sec:PIMD_result}}
\begin{figure}[htbp]
	\centering
	\includegraphics[width=0.5\textwidth]{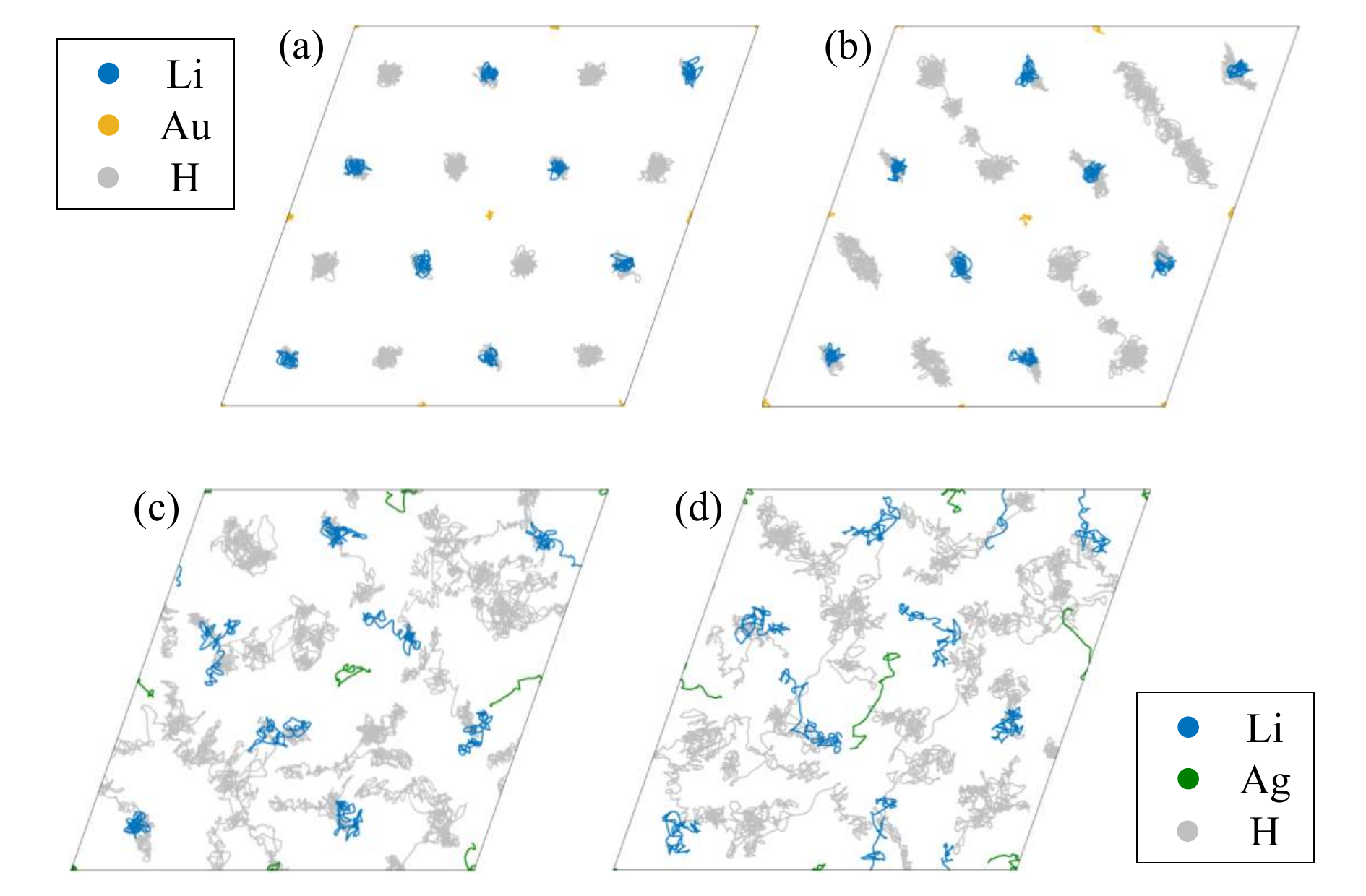}
	\caption{[100] view of trajectories of centroid mode of all lithium (blue), gold (yellow), and hydrogen (gray) atoms in the 1-ps (a) MD and (b) PIMD simulation for Li$_2$AuH$_6$ in a $2\times2\times2$ supercell at 80~K. (c) and (d) are the corresponding MD and PIMD trajectories for Li$_2$AgH$_6$, with silver atoms marked by green points.}
	\label{fig:PIMD_sc72}
\end{figure}
The results of MD and PIMD simulations in the $2\times2\times2$ supercell are shown in Fig.~\ref{fig:PIMD_sc72}. Figures~\ref{fig:PIMD_sc72}(a) and \ref{fig:PIMD_sc72}(b) illustrate the trajectories of all atoms in Li$_2$AuH$_6$ in MD and PIMD simulations, respectively. In the MD simulation, all atoms vibrate near their equilibrium positions without diffusion, which suggests a solid state. However, in the PIMD simulation, although lithium and gold atoms still vibrate near their equilibrium positions, hydrogen atoms begin to move away from the equilibrium positions of the solid state, which resembles a superionic state. This suggests that the quantum fluctuations of hydrogen atoms induce kinetic instability in Li$_2$AuH$_6$.

Figures~\ref{fig:PIMD_sc72}(c) and \ref{fig:PIMD_sc72}(d) illustrate the trajectories of all atoms in Li$_2$AgH$_6$ in MD and PIMD simulations, respectively. It can be seen that all atoms begin to move away from their initial positions in both MD and PIMD simulations and the $Fm\bar{3}m$ crystal structure collapses rapidly. This indicates that Li$_2$AgH$_6$ undergoes lattice destabilization and thus is not kinetically stable.

\begin{figure}[htbp]
	\centering
	\includegraphics[width=0.5\textwidth]{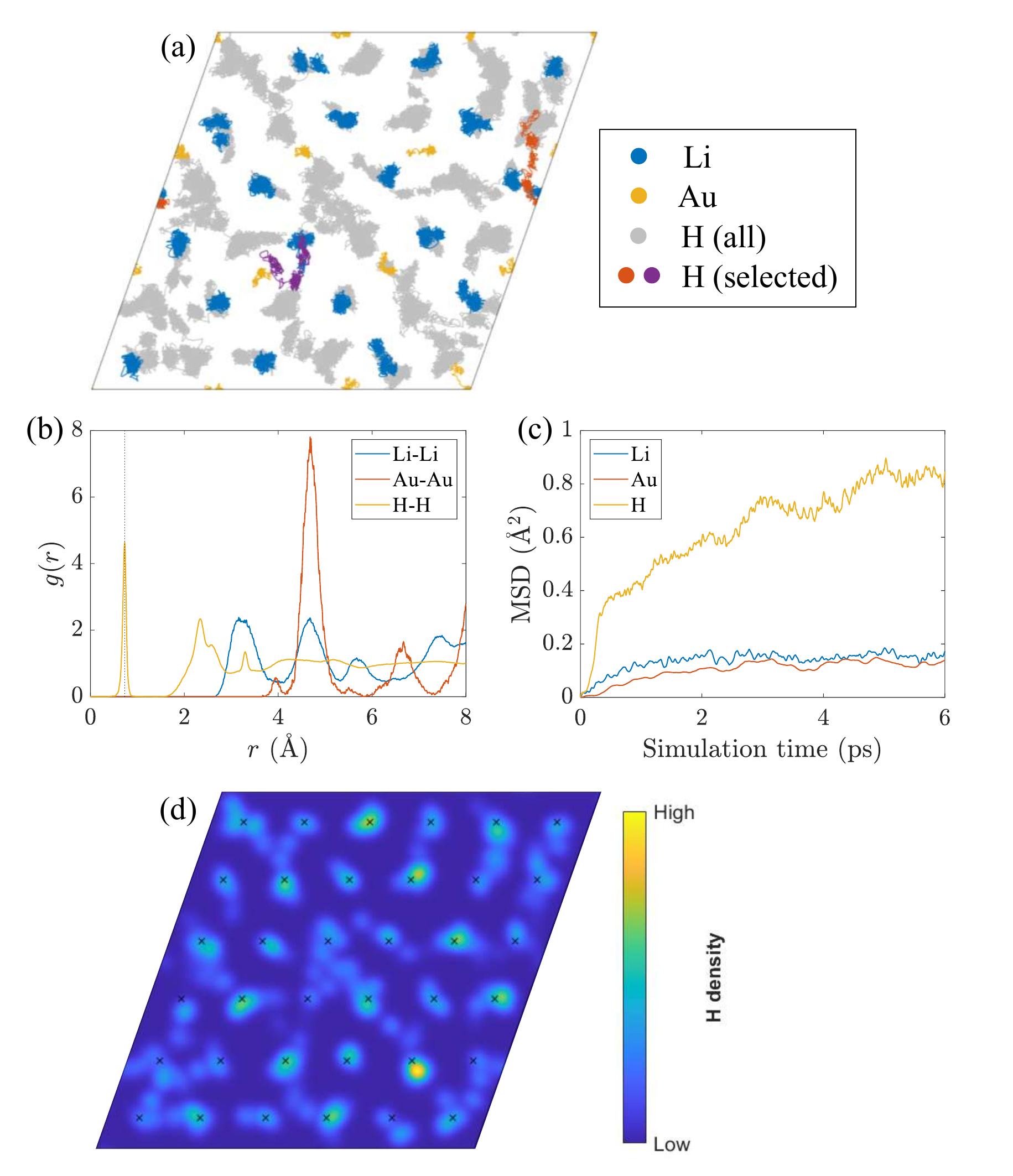}
	\caption{(a) [100] view of trajectories of centroid mode of all lithium (blue), gold (yellow), and hydrogen (gray) atoms in the 6-ps PIMD simulation for Li$_2$AuH$_6$ in a $3\times3\times3$ supercell at 80~K. The trajectories of two selected hydrogen atoms with large displacements are marked by orange and purple points. (b) and (c) are the corresponding RDF and MSD. The black dotted line in (b) marks the first peak of the H-H RDF at 0.74~\AA. (d) [100] view of the density distribution of hydrogen atoms during the PIMD simulation. Hydrogen atoms of the original solid structure are marked with cross marks.}
	\label{fig:PIMD_sc243}
\end{figure}
Figure~\ref{fig:PIMD_sc243} shows the results of a longer PIMD simulation for Li$_2$AuH$_6$ in a larger supercell. Figure~\ref{fig:PIMD_sc243}(a) illustrates the trajectories of all atoms, with trajectories of selected hydrogen atoms highlighted. Their radial distribution functions (RDF) and mean squared displacements (MSD) are shown in Figs.~\ref{fig:PIMD_sc243}(b) and \ref{fig:PIMD_sc243}(c), respectively. It can be seen that lithium and gold atoms exhibit large vibration amplitudes around their equilibrium positions, but they still maintain the fluorite-type crystal structure. The hydrogen atoms, however, begin to slowly diffuse in the system without equilibrium positions, as shown by the trajectories of the selected atoms and the increasing MSD. These results confirm that Li$_2$AuH$_6$ is not kinetically stable, albeit with a stable Li-Au host framework.

Notably, the hydrogen atoms partially dimerize and form hydrogen molecules in Li$_2$AuH$_6$. As shown in Fig.~\ref{fig:PIMD_sc243}(b), the H-H RDF exhibits a sharp peak near 0.74~\AA, which is exactly the bond length of the hydrogen molecule at ambient pressure. We evaluate the proportion of hydrogen atoms with a neighboring hydrogen atom within 1~\AA\ during the PIMD simulation and find that 23.6\% of hydrogen atoms dimerize into molecules. The hydrogen dimers appear to be stable, since each hydrogen atom within a dimer maintains the same nearest neighbor throughout the simulation.

To further analyze the rearranged hydrogen structure, we calculate the hydrogen density distribution by representing each hydrogen atom with a 3D Gaussian and accumulating over all PIMD configurations. As shown in Fig.~\ref{fig:PIMD_sc243}(d), the hydrogen density distribution deviates significantly from that of the original solid structure. Although the hydrogen density remains high near most of the solid equilibrium positions, it also populates sites that are unoccupied in the initial crystal structure. The hydrogen density distribution is delocalized and spread out, exhibiting no distinct equilibrium positions, so the hydrogen diffusion is likely to persist. These results contrast with prior beliefs that Li$_2$AuH$_6$ has a metastable solid structure with atomic hydrogens. 

\section{Superconductivity of L\lowercase{i}$_2$A\lowercase{u}H$_6$\label{sec:SC}}
\subsection{Stochastic path-integral approach\label{sec:SPIA}}
For the Li$_2$AuH$_6$ system with diffusive hydrogen atoms, conventional approaches such as density functional perturbation theory (DFPT)~\cite{DFPT} and SSCHA, which assume the existence of atomic equilibrium positions, clearly do not apply. On the other hand, SPIA is a nonperturbative method for determining $T_c$ of conventional superconductors based on PIMD~\cite{SPIA}. It does not make assumptions about the nature of ion motion and is thus applicable to our system. It has been successfully applied not only to solid systems~\cite{H3S_SPIA,YH_SPIA}, but also to systems with diffusive atoms~\cite{H_SPIA,LMH_SPIA,LaH10_SPIA}.

In SPIA, we calculate the fluctuation of the electron-ion scattering $T$ matrix in PIMD simulations, i.e., $\Gamma_{11^\prime}=-\beta\langle\mathcal{T}_{11^\prime}[\boldsymbol{R}(\tau)]\mathcal{T}_{\bar{1}\bar{1}^\prime}[\boldsymbol{R}(\tau)]\rangle_C$, where $\hat{\mathcal{T}}[\boldsymbol{R}(\tau)]$ is the electron-ion scattering $T$ matrix with respect to the imaginary time-dependent ion configuration $\boldsymbol{R}(\tau)$, and $\langle\cdots\rangle_C$ denotes the average over PIMD sampling configurations. $\Gamma_{11^\prime}$ describes the scattering amplitude of a time-reversal electron pair from ($1,\bar{1}$) to ($1^\prime,\bar{1}^\prime$). The effective electron-electron interaction $\hat{W}$ induced by ion motion is determined by solving the Bethe-Salpeter equation
\begin{equation}
	W_{11^\prime}=\Gamma_{11^\prime}+\frac{1}{\hbar^2\beta}\sum_2W_{12}|\bar{\mathcal{G}}_2|^2\Gamma_{21^\prime},
\end{equation}
where $1=(n,\boldsymbol{k},\omega_j)$ is the state index for the generalized Bloch states that diagonalize the electron Green's function $\hat{\bar{\mathcal{G}}}$~\cite{H3S_SPIA,LMH_SPIA}, with $n$ and $\boldsymbol{k}$ being band and quasi-wave vector indices, and $\omega_j$ being the Fermionic Matsubara frequency. As rigorously established in Ref.~\cite{SPIA}, the effective electron-electron interaction so determined enters the linearized Eliashberg equation~\cite{Allen-Dynes} in a manner analogous to conventional theories for determining $T_c$.

For Li$_2$AuH$_6$, we apply the isotropic approximation~\cite{ambient_maximum} and calculate the EPC parameters $\lambda_0(j-j^\prime)$ at the simulation temperature $T_0$ by averaging the effective interaction $\hat{W}$ over the Fermi surface, i.e.,
\begin{equation}
	\lambda_0(j-j^\prime)=-N(\epsilon_F)\langle W_{11^{\prime}}\rangle_{\mathrm{FS}},
\end{equation}
where $j-j^\prime$ gives a bosonic Matsubara frequency $\nu_{j-j^\prime}=\omega_j-\omega_{j^\prime}$, $N(\epsilon_F)$ is the electron DOS at the Fermi level determined by averaging the DOS over all PIMD configurations, and $\langle\cdots\rangle_{\mathrm{FS}}$ denotes the average for all initial states $(n\boldsymbol{k})$ and final states $(n^\prime\boldsymbol{k}^\prime)$ over the Fermi surface. We then define a continuous frequency-dependent function $\Lambda(\nu)$ by interpolating the set of EPC parameters, with the assumption $\lambda_0(m)\equiv\Lambda(2\pi mk_BT_0/\hbar)$ (see Fig.~\ref{fig:lambda}(a)). EPC parameters at other temperatures are inferred from $\Lambda(\nu)$ with $\lambda(m)=\Lambda(2\pi mk_BT/\hbar)$~\cite{H3S_SPIA}.

\subsection{Computational details\label{sec:SPIA_computation}}
The SPIA calculation of Li$_2$AuH$_6$ is based on the 6-ps PIMD simulation in the $3\times3\times3$ supercell. The ion configurations of Li$_2$AuH$_6$ are uniformly sampled with a spacing of 40 time steps. We find that a simulation time of 1~ps after a 1-ps equilibration is sufficient to obtain converged results. DFT calculations are performed for these configurations. An energy cutoff of 225~eV for plane waves is used to expand electron wave functions, and a $2\times2\times2$ $\Gamma$-centered $\boldsymbol{k}$-point grid is used to sample the Brillouin zone of the supercell. The converged DFT results are then used as inputs of our MATLAB implementation of SPIA~\cite{SPIA_program}. Finally, the effective electron-electron interaction $\hat{W}$ is calculated on a $3\times3\times3$ $\boldsymbol{k}$-point grid of the supercell. When averaging $\hat{W}$ near the Fermi surface to obtain EPC parameters, a Lorentzian smearing is used~\cite{SPIA} with a half-width of 0.1~eV. The Morel-Anderson pseudopotential~\cite{mustar} is set to a typical value $\mu^*=0.1$~\cite{Li2AuH6}. The convergence of calculation is tested in the Appendix.

\subsection{SPIA results\label{sec:SPIA_result}}
\begin{figure}[htbp]
	\centering
	\includegraphics[width=0.5\textwidth]{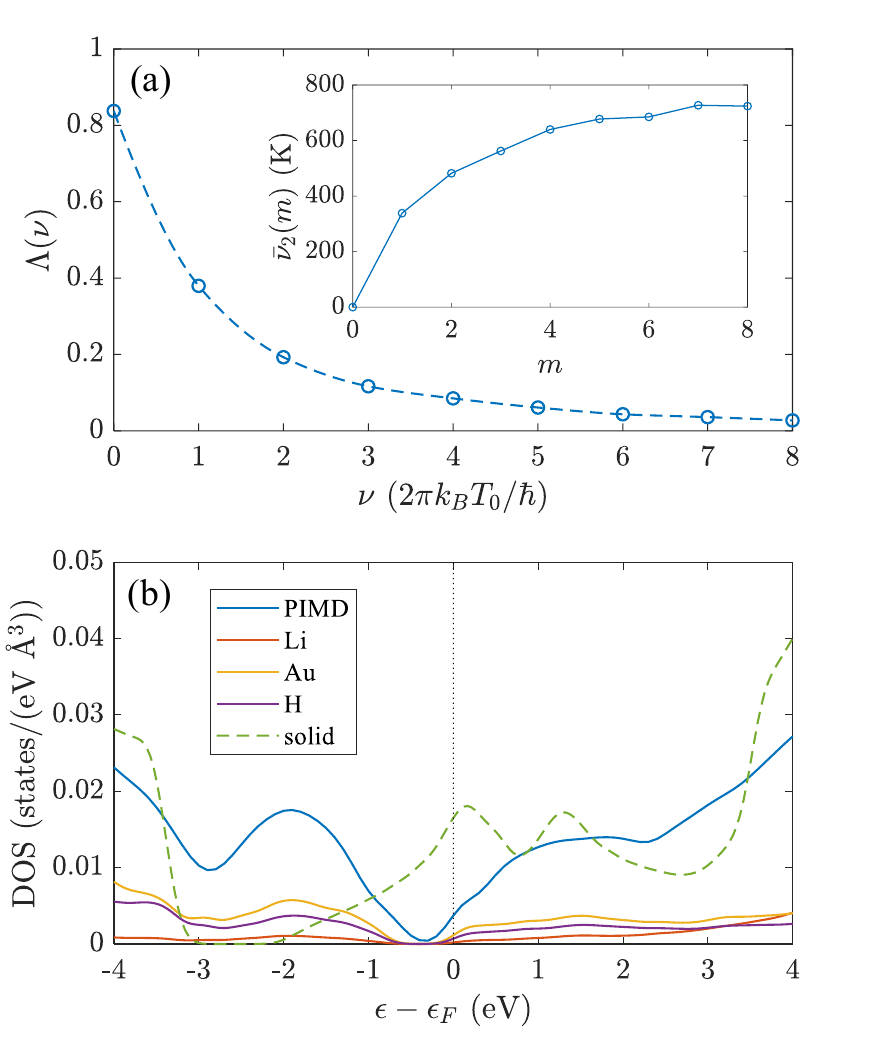}
	\caption{(a) The $\Lambda(\nu)$ curve of Li$_2$AuH$_6$ (dashed line) interpolated from its EPC parameters $\lambda_0(m)\equiv\Lambda(2\pi mk_BT_0/\hbar)$ at the simulation temperature $T_0=80$~K (circles). The inset shows the asymptotic behavior of $\bar{\nu}_2(m)=2\pi/\hbar\beta\sqrt{m^2\lambda_{0}(m)/\lambda_{0}(0)}$. (b) The DOS of Li$_2$AuH$_6$ calculated by averaging the DOS over all PIMD configurations (blue line). The Fermi level is marked by the black dotted line. The red, yellow, and purple lines show the atom-projected DOS of one PIMD configuration using PAW projections~\cite{PAW}. The green dashed line denotes the DOS of the initial solid structure.}
	\label{fig:lambda}
\end{figure}

Figure~\ref{fig:lambda}(a) shows the interpolated $\Lambda(\nu)$ curve of Li$_2$AuH$_6$ and the corresponding EPC parameters $\lambda_{0}(m)$ at 80~K. It also illustrates the behavior of $\bar{\nu}_2(m)=2\pi/\hbar\beta\sqrt{m^2\lambda_{0}(m)/\lambda_{0}(0)}$, whose asymptotic value gives the average phonon frequency $\bar{\nu}_2$. The EPC parameter $\lambda(0)=0.838$ is much smaller than the typical values from 2.10 to 2.84 in previous studies, while the average phonon frequency $\bar{\nu}_2=758$~K is close to prior results from 724~K to 982~K~\cite{Li2AuH6,ambient_maximum,Li2AuH6_pressure}. The resulting $T_c$ is 22~K by solving the linearized Eliashberg equation, also well below prior predictions of 80--140~K.

The considerable depression of EPC parameters and the predicted $T_c$ mainly results from the decrease in DOS at the Fermi level. Figure~\ref{fig:lambda}(b) shows the DOS calculated from all PIMD configurations and the initial solid structure, as well as the atom-projected DOS of one PIMD configuration. The atom-projected DOS of the solid structure can be found in Ref.~\cite{Li2AuH6}. It can be seen that the DOS obtained using PIMD configurations exhibits a dip near the Fermi level, with $N(\epsilon_F)$ being only 0.0038~states/(eV~\AA$^3$), in sharp contrast to that of the solid structure where a pronounced peak appears near the Fermi level~\cite{Li2AuH6}. The atom-projected DOS further reveals that the electronic states contributed by hydrogen atoms, which dominate the DOS peak at the Fermi level for the solid structure~\cite{Li2AuH6}, are now strongly suppressed. This can be attributed to the collapse of the hydrogen sublattice and the dimerization of hydrogen atoms into molecules (see Figs.~\ref{fig:PIMD_sc243}(b) and \ref{fig:PIMD_sc243}(d)).

In our calculations, we assume the $\Lambda(\nu)$ curve, which characterizes the EPC, is universal for all temperatures. This implies the properties of ion motion do not change at different temperatures~\cite{H3S_SPIA}. However, since our PIMD simulations are performed at 80~K, which is chosen based on previous $T_c$ predictions, and the resulting $T_c$ is considerably lower (22~K), this assumption may not hold. Nevertheless, it at least demonstrates that Li$_2$AuH$_6$ has a $T_c$ far below 80~K, as the collapse of hydrogen sublattice, accompanied by hydrogen diffusion and dimerization, already occurs at temperatures well below 80~K (see Fig.~\ref{fig:PIMD_50K}). Besides, the Morel-Anderson pseudopotential could be larger than 0.1 for molecular hydrogen systems~\cite{H_atomic,H_molecular}, which would further decrease the predicted $T_c$. These all suggest that Li$_2$AuH$_6$ is unlikely to be a high-$T_c$ superconductor under ambient pressure.

\section{Summary\label{sec:summary}}
In conclusion, we study the kinetic stability and superconductivity in Li$_2$AuH$_6$ and Li$_2$AgH$_6$ at ambient pressure. We find both compounds are kinetically unstable. Li$_2$AgH$_6$ undergoes lattice collapse, whereas Li$_2$AuH$_6$ retains a stable Li-Au host lattice with hydrogen atoms diffusing and partially dimerizing into molecules. Moreover, the predicted $T_c$ for Li$_2$AuH$_6$ is significantly suppressed compared to previous studies due to the low DOS at the Fermi level. We thus conclude that both Li$_2$AuH$_6$ and Li$_2$AgH$_6$ are unlikely to be high-$T_c$ superconductors under ambient pressure. 

\begin{acknowledgments}
We thank Yuewen Fang for useful suggestions. This work is supported by the National Natural Science Foundation of China under Grant Nos.~12174005 and 12574169, and the National Key R\&D Program of China under Grant No.~2021YFA1401900.
\end{acknowledgments}

\appendix*
\section{Convergence of calculation\label{sec:convergence}}
\begin{figure}[htbp]
	\centering
	\includegraphics[width=0.5\textwidth]{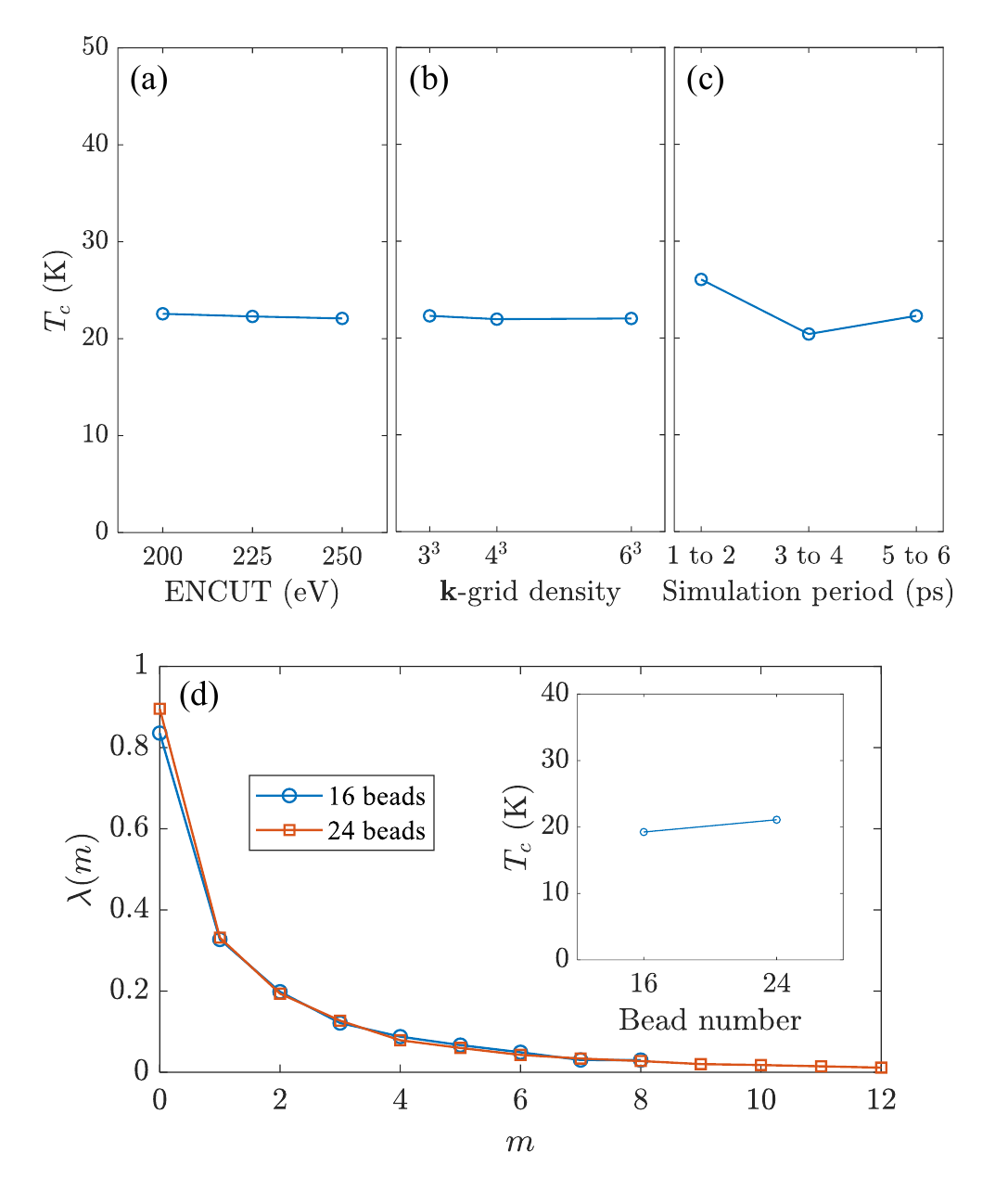}
	\caption{$T_c$ values of Li$_2$AuH$_6$ for (a) varying plane-wave energy cutoffs, (b) the $\boldsymbol{k}$-grid densities of the supercell, and (c) different simulation time periods (see Fig.~\ref{fig:PIMD_sc243}(c)) used for $T_{c}$ estimation. (d) EPC parameters and $T_c$ for different PIMD bead numbers in a $2\times2\times2$ supercell.}
	\label{fig:convergence}
\end{figure}

In this section, we first test the convergence of our SPIA calculation with respect to the plane-wave energy cutoff, the $\boldsymbol{k}$-grid density of the supercell, and the PIMD simulation period. As shown in Figs.~\ref{fig:convergence}(a) to \ref{fig:convergence}(c), a plane-wave energy cutoff of 225~eV, a $3\times3\times3$ $\boldsymbol{k}$-point grid of the supercell, and a 1~ps simulation after equilibrium is sufficient to obtain a converged $T_c$ for Li$_2$AuH$_6$.

To test the convergence with respect to bead numbers in PIMD, we perform PIMD and SPIA calculations for Li$_2$AuH$_6$ in a smaller $2\times2\times2$ supercell using 16 and 24 beads, respectively. As shown in Fig.~\ref{fig:convergence}(d), the resulting EPC parameters are close, and the $T_c$ values differ by approximately 1~K. This indicates that a bead number of 16 is sufficient for achieving convergence in our calculations.

\begin{figure}[htbp]
	\centering
	\includegraphics[width=0.5\textwidth]{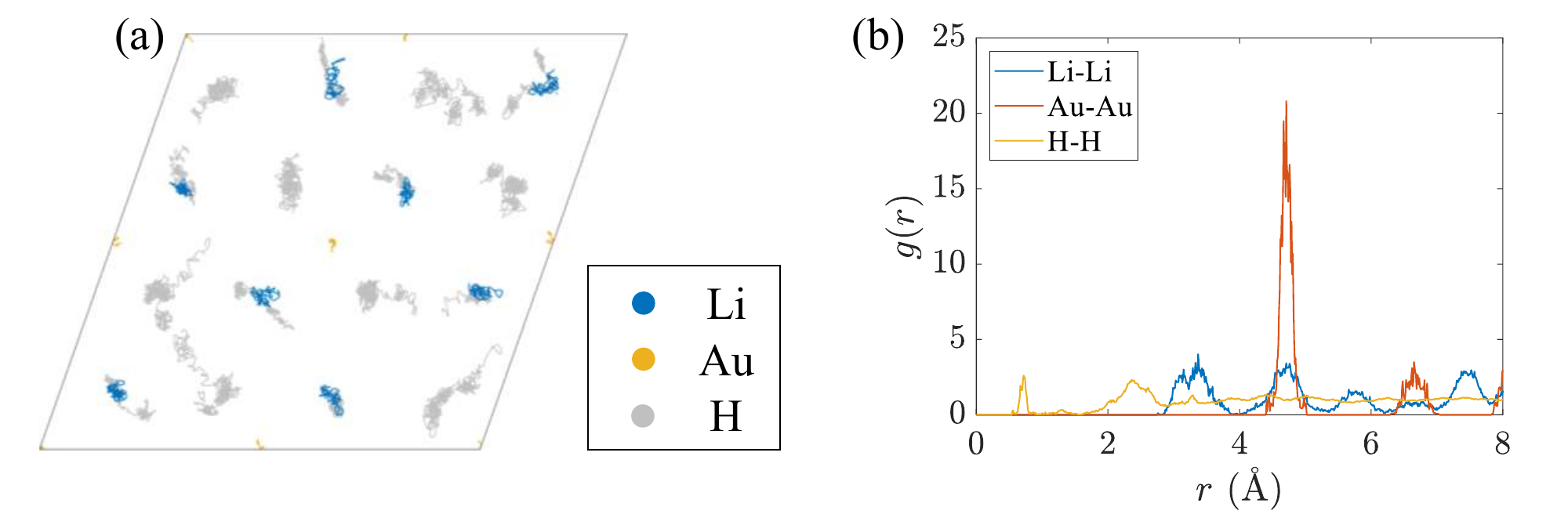}
	\caption{Results of a 1-ps PIMD simulation for Li$_2$AuH$_6$ in the $2\times2\times2$ supercell with 16 beads at 50~K. (a) [100] view of trajectories of centroid mode of all lithium (blue), gold (yellow), and hydrogen (gray) atoms. (b) is the corresponding RDF.}
	\label{fig:PIMD_50K}
\end{figure}

To explore the physical state of Li$_2$AuH$_6$ at lower temperatures, we perform a PIMD simulation in the $2\times2\times2$ supercell using 16 beads at 50~K. As shown in Fig.~\ref{fig:PIMD_50K}, the same hydrogen sublattice collapse, hydrogen diffusion, and hydrogen dimerization are clearly observed. These structural features lead to a low DOS at the Fermi level, consistent with our prediction of a $T_c$ well below 80~K.

\nocite{*}
\bibliography{reference}

\end{document}